# Gap state analysis in electric-field-induced band gap for bilayer graphene

Kaoru Kanayama[1], and Kosuke Nagashio[1,2*]
[1]Department of Materials Engineering, The University of Tokyo, Tokyo 113-8656, Japan
[2]PRESTO, Japan Science and Technology Agency (JST), Tokyo 113-8656, Japan
[*]nagashio@material.t.u-tokyo.ac.jp

**The origin of the low current on/off ratio at room temperature in dual-gated bilayer graphene field-effect transistors is considered to be the variable range hopping in gap states. However, the quantitative estimation of gap states has not been conducted. Here, we report the systematic estimation of the energy gap by both quantum capacitance and transport measurements and the density of states for gap states by the conductance method. An energy gap of ~250 meV is obtained at the maximum displacement field of ~3.1 V/nm, where the current on/off ratio of ~3×10$^3$ is demonstrated at 20 K. The density of states for the gap states are in the range from the latter half of 10$^{12}$ to 10$^{13}$ eV$^{-1}$cm$^{-2}$. Although the large amount of gap states at the interface of high-*k* oxide/bilayer graphene limits the current on/off ratio at present, our results suggest that the reduction of gap states below ~10$^{11}$ eV$^{-1}$cm$^{-2}$ by continual improvement of the gate stack makes bilayer graphene a promising candidate for future nanoelectronic device applications.**

The main issue of downscaling in the Si field-effect transistors (FETs) is the short channel effect in which the gate control is weakened by the drain bias. Based on an analysis of the electrical potential distribution in the channel region, it is well known that the short channel effect can be neglected when the channel length is ~6 times longer than the scaling length $\lambda = \sqrt{(\varepsilon_{ch} t_{ch} t_{ox})/(N \varepsilon_{ox})}$,[1,2] where $N$, $\varepsilon_{ch}$, $\varepsilon_{ox}$, $t_{ch}$, and $t_{ox}$ are the effective gate number, dielectric constants for the channel and gate insulator, and thickness of the channel and the gate oxide. This perspective attracts great attention to two-dimensional (2D) layered channels in the FET application because of their rigidly controllable atomic thickness ($t_{ch}$< 1 nm), as well as the low dielectric constant ($\varepsilon_{ch}$ = ~4) for typical 2D layered channels[3,4]. Although old-but-new 2D channels, such as transition metal dichalcogenides, black phosphorus, and so on, have been intensively investigated recently[5-9], bilayer graphene with an electrostatically tunable band gap still has an advantage over the high performance device from the viewpoint of mobility due to the smaller effective mass ($m_{BLG}$=~0.037[10], $m_{BP}$=0.13[11], and $m_{MoS2}$=~0.37[12]).

The suppression of conductivity in bilayer graphene has so far been reported by many researchers by applying an external electrical field[13-24]. Optical spectroscopic measurements, such as angle-resolved photoemission spectroscopy[25] and infrared spectroscopy[14,26,27], confirmed the band gap formation. However, the large current on/off ratio ($I_{on}/I_{off}$) of ~10$^6$ is obtained only at the quite low temperature of 300 mK[17], not at room temperature. The reason is explained by the variable range hopping in gap states[13,15,17-19]. Therefore, the main target issue for bilayer graphene is a low current on/off ratio at room temperature. At the zero-order approximation, there will intrinsically be no interface states in bilayer graphene because there are no dangling bonds on the basal plane, compared with $P_b$ centers in the $SiO_2$/Si system, assigned by the electron spin resonance measurement[28-30]. Although the strong disorder at the channel edge was expected to act as a main conduction path, the transport measurement in the Corbino geometry excluded this idea[19]. The origin of the gap states remains an open question. So far, detailed measurements on the density of states ($D_{it}$) and the time constant ($\tau_{it}$) for gap states have not been reported.

Contrary to the transport measurement, the extraction of the quantum capacitance ($C_Q$) through the capacitance measurements ($C$-$V$) of bilayer graphene provides direct information on the density of states ($DOS$) of bilayer graphene, consequently, the energy gap ($E_G$), because it is regarded as the energy cost of inducing the carriers in graphene and is directly related as $C_Q = e^2DOS$[31]. Although there are a few reports on $C_Q$ measurement for bilayer graphene[32,33,34], the comparison of $E_G$ estimated both from $I$-$V$ and $C$-$V$ has not been done yet. Moreover, in principle, the mobile carrier response to a small-signal alternating current voltage at a certain frequency is measured in the $C$-$V$ measurement. The capture and emission process of mobile carriers at the trap levels distributed throughout the band gap can be extracted as a deviation from the ideal carrier response without any trap levels under the assumption of the equivalent circuit. This technique is known as the conductance method[34].

In this work, we present the systematic extraction of $E_G$ as a function of the displacement field ($\overline{\mathrm{D}}$), which determines the band structure of bilayer graphene, from both $I$-$V$ and $C$-$V$. The conductance measurements are carried out to extract $D_{it}$ and $\tau_{it}$; then, the possible origins of the gap states are discussed.

## Bilayer graphene FTEs with a high quality top gate insulator

Recently, we have demonstrated a considerable suppression of the low-field leakage through high-$k$ $Y_2O_3$ on monolayer graphene by applying high-pressure $O_2$ annealing[36]. The same process was applied to bilayer graphene FETs. The improved electrical quality of the insulators provides access to the large displacement field ($\overline{\mathrm{D}}$ = ~3.1 V/nm) in this study. Although there are several conventions for $\overline{\mathrm{D}}$, we adopt the most widely used definition of $\bar{D} = 1/2\left[\varepsilon_{BG}/d_{BG}\left(V_{BG}-V_{BG}^0\right)-\varepsilon_{TG}/d_{TG}\left(V_{TG}-V_{TG}^0\right)\right]$ in this study[14], where $\varepsilon_{BG}$, $\varepsilon_{TG}$, $d_{BG}$, $d_{TG}$, $V_{BG}$, and $V_{TG}$ are the dielectric constants, the insulator thickness, and the gate voltages for back- and top-gate insulators, respectively. $(V_{TG}^0, V_{BG}^0)$ is the charge neutrality point to give the minimum resistance in the top-gated region.

To suppress the hysteresis in drain current - gate voltage curves, the $SiO_2$ surface was converted to be hydrophobic (siloxane group) by annealing the $SiO_2$/Si substrate in a 100 % $O_2$ atmosphere at 1000 °C prior to the graphene transfer[37]. Then, the conventional back-gated bilayer graphene FETs containing source and drain electrodes were fabricated on ~90 nm $SiO_2$/n$^+$-Si substrates by the mechanical exfoliation of Kish graphite. This device was annealed under Ar/$H_2$ gas flow at 300 °C for 3 hours to remove the resist residue on the bilayer graphene channel. Subsequently, $Y_2O_3$, with a thickness of ~6 nm, was deposited on bilayer graphene FETs by the thermal evaporation of Y metal in the $O_2$ atmosphere at $P_{O2} = 10^{-1}$ Pa[36]. Then, high-pressure annealing was carried out in a 100 % $O_2$ atmosphere at ~100 atm and 300 °C. Finally, the top gate electrode was patterned, followed by annealing at 300 °C for 30 s under 0.1% $O_2$ gas flow. The lack of a Raman D band measured through $Y_2O_3$ indicated that no detectable defects were introduced into the bilayer graphene by the high-pressure $O_2$ annealing, as shown in **supplementary Fig. S1c**.

## Estimation of $E_G$ by $C$-$V$ and $I$-$V$

We first focus on the capacitance measurement to determine $E_G$ through the $C_Q$ extraction. **Figure 1a** shows the total capacitance ($C_{Total}$) between the source and top gate electrodes, obtained by sweeping $V_{TG}$ at different $V_{BG}$, which were measured at the frequency of 1 MHz in a vacuum of ~$1\times10^{-5}$ Pa at 20 K. **Figure 1b** is the counter plot of $C_{Total}$. The $C_{Total}$ reduction at the Dirac point with increasing $V_{BG}$ indicates the decrease in the $DOS$ by the gap formation because of the strong contribution of $C_Q$. The hysteresis in the bidirectional $C$-$V$ curves is quite small (~0.1 V for a $V_{TG}$ = ±4 V sweep).[36] As shown in **supplementary Fig. S2**, the frequency dependence of $C_{Total}$ is clearly observed in the gap region, suggesting the existence of trap sites. The gradual saturation of the capacitance with increasing frequency from 200 kHz to 1 MHz suggests that $C_{Total}$ measured at 1 MHz is close to the ideal capacitance without any response to the trap site. Therefore, the $C$-$V$ measurement in **Fig. 1a** was carried out at 1 MHz.

The slope of the dotted black line at the charge neutrality point of $\left(V_{TG}^0, V_{BG}^0\right) = (0.75, 9)$ in **Fig. 1b** corresponds to the capacitive coupling ratio between the top gate and back gate, that is, –

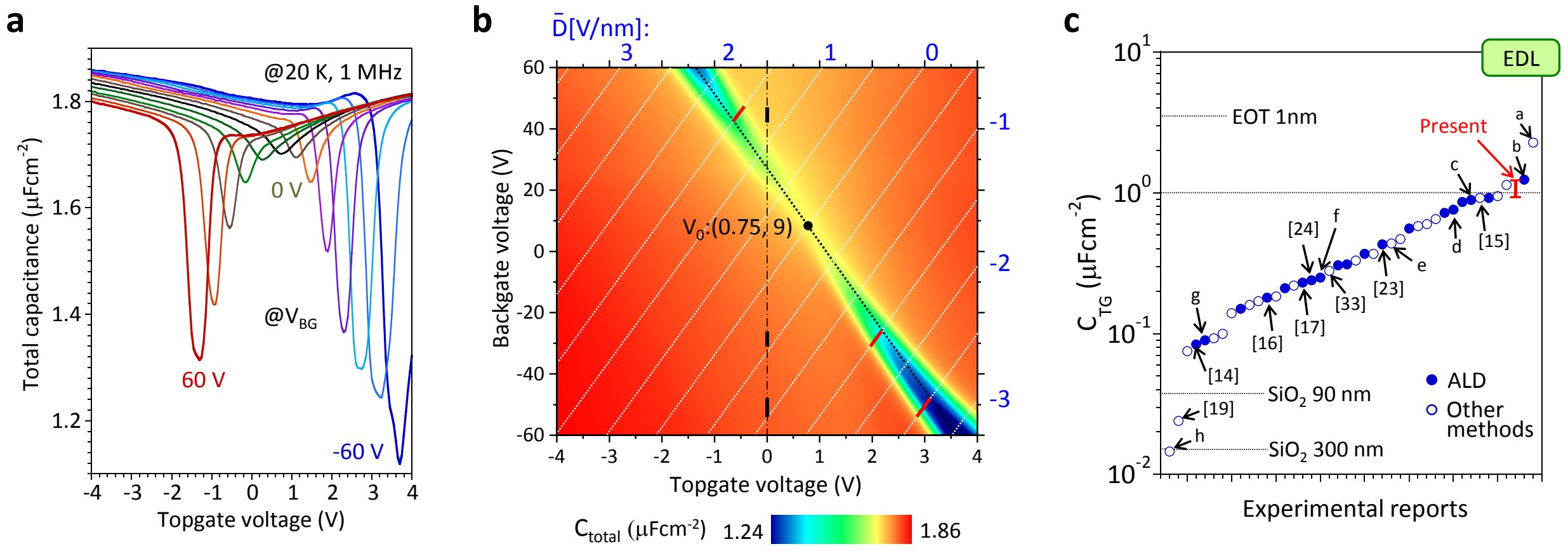

**Figure 1 | Capacitance characteristics of dual gated bilayer graphene FETs.**

(a) $C_{\mathrm{Total}}$ between the source and top gate electrodes as a function of $V_{\mathrm{TG}}$ for different $V_{\mathrm{BG}}$ measured at 20 K and a frequency of 1 MHz. (b) Counter plot of $C_{\mathrm{Total}}$. The charge neutrality point is $(V_{TG}^0, V_{BG}^0) = (0.75, 9)$. The $\bar{\mathrm{D}}$ value is shown at the periphery of the counter plot. (c) Comparison of $C_{\mathrm{TG}}$ with those reported previously for both monolayer and bilayer graphene. Closed and open circles indicate $C_{\mathrm{TG}}$ obtained for oxide insulators deposited by the atomic layer deposition technique (ALD) and for insulators prepared by another technique, respectively. "EOT" and "EDL" indicate $C_{\mathrm{TG}}$ obtained for 1 nm-thick-$SiO_2$ and a typical electric double-layer, respectively. Several papers are included. a: Zhang, Z. *et al. Appl. Phys. Lett.* **101,** 213103 (2012). b: Fallahazad, B. *et al. Appl. Phys. Lett.* **100,** 093112 (2012). c: Zou, K. *et al. Nano Lett.* **13,** 369 (2013). d: Meric, I. *et al. Nature Nanotech.* **3,** 654 (2008). e: Liao, L. *et al., Nature* **467,** 305 (2010). f: Kim, S. *et al. Appl. Phys. Lett.* **94**, 062107 (2009). g: Wu, Y. *et al. Nature* **472,** 74 (2011). h: Velasco Jr., J. *et al. Nature Nanotech.* **7,** 156 (2012).

$0.0412 = -C_{\mathrm{BG}}/C_{\mathrm{TG}}$. Using $C_{\mathrm{BG}}$=0.038 μFcm$^{-2}$ for $SiO_2$ with $d_{\mathrm{BG}}$ = 90 nm and $\varepsilon_{\mathrm{BG}}$ = 3.9, $C_{\mathrm{TG}}$ is estimated to be 0.93 μFcm$^{-2}$. In the course of this study, the typical $C_{\mathrm{TG}}$ value is ~1.2 μFcm$^{-2}$, depending on the $Y_2O_3$ thickness. The $C_{\mathrm{TG}}$ value is considerably high compared with those reported previously for both monolayer and bilayer graphene, as shown in **Fig. 1c**. Based on $C_{\mathrm{TG}}$ and $C_{\mathrm{BG}}$, the white dotted lines indicate the constant $\bar{\mathrm{D}}$, whose values are shown at the periphery of the counter plot of **Fig. 1b**. The maximum $\bar{\mathrm{D}}$ at the Dirac point in this study is ~3.1 V/nm, which is quite high compared with other reports.

The simplified equivalent circuit model of the device is shown in **Fig. 2a**, where $V_{\mathrm{ch}}$ and $C_{\mathrm{para}}$ are the charging voltage and the parasitic capacitance. The Fermi energy ($E_{\mathrm{F}}$) and the band structure of bilayer graphene are independently controlled by changing $V_{\mathrm{TG}}$ and $V_{\mathrm{BG}}$. It should be noted that the contribution of $C_{\mathrm{BG}}$ is implicitly involved in $C_{\mathrm{Total}}$ through $C_{\mathrm{Q}}$ and $V_{\mathrm{ch}}$, in **Fig. 2a**. Based on the equivalent circuit of **Fig. 1a**, $C_{\mathrm{Q}}$ was extracted along the constant $\bar{\mathrm{D}}$ lines, i.e., the constant band structure, by using $C_{\mathrm{para}}$ as a fitting parameter. **Figure 2b** shows $C_{\mathrm{Q}}$ as a function of $E_{\mathrm{F}}$ for different constant $\bar{\mathrm{D}}$ values. At $\bar{\mathrm{D}}$ = 0 V/nm, the extracted $C_{\mathrm{Q}}$ is fitted reasonably well with the theoretical value for bilayer graphene calculated by the tight-binding model[38] by selecting $C_{\mathrm{para}}$ = 0.91 μFcm$^{-2}$. The charging energy required to induce carriers in bilayer graphene is denoted by $E_{\mathrm{F}}$, which is expressed as $E_{\mathrm{F}} = eV_{\mathrm{ch}}$. $V_{ch}$ can be expressed by the equation for a series combination of capacitors according to $V_{ch} = V'_{TG} - \int_0^{V'_{TG}} C'_{Total}/C_{Y2O3}\, dV'_{TG}$, where $C'_{\mathrm{Total}}$ and $V'_{TG}$ are defined as $C'_{\mathrm{Total}} = C_{\mathrm{Total}} - C_{\mathrm{para}}$ and $V'_{TG} = V_{TG} - V_{DP}$, respectively. $V_{\mathrm{DP}}$ is the Dirac point voltage. The detailed calculation method is explained in **supplementary Fig. S3**. The $C_{\mathrm{Q}}$ value at the Dirac point for $\bar{\mathrm{D}}$ = 0 V/nm is consistent well with the theoretical value because the *DOS* for bilayer graphene at the Dirac point is larger than the residual carriers induced externally by the charged impurities at the $SiO_2$ surface.[39] With increasing $\bar{\mathrm{D}}$, the reduction of $C_{\mathrm{Q}}$ is clearly observed because the scattering issue, which strongly contributes to the conductivity in *I*-*V*, can be neglected in *C*-*V*. It should be emphasized that the *DOS* within the gap region almost reaches zero, which is not observed in the previous report

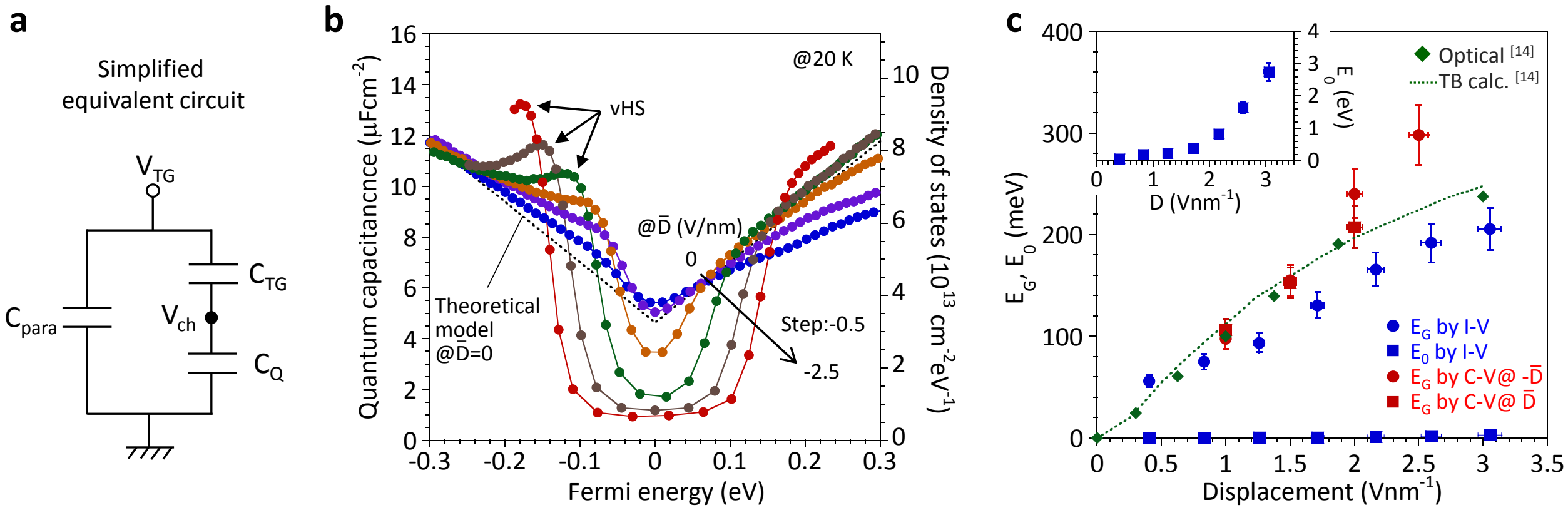

**Figure 2 | Quantum capacitance and energy band gap.**

(a) Simplified equivalent circuit for the bilayer graphene FET. (b) $C_Q$ as a function of $E_F$ at different $\bar{D}$ ranging from ~0 to −2.5 V/nm. The right vertical axis indicates *DOS*. (c) $E_G$ and $E_0$ as a function of absolute value for $\bar{D}$ determined from the *C-V* data and *I-V* data.

for bilayer graphene with the *h*-BN top gate[33]. Moreover, the van Hove Singularity (vHS) is also observed near the valence band edge, as shown by arrows. The appearance of the vHS is asymmetric, that is, a valence band edge for negative $\bar{D}$ and a conduction band edge for positive $\bar{D}$ (not shown in **Fig. 2b**), which is consistent with previous data[33]. This phenomenon is explained by the near-layer capacitance enhancement effect[40]. **Figure 2c** shows $E_G$ as a function of absolute value for $\bar{D}$, determined from the *C-V* data. $E_G$ was defined as the energy between inflection points for the conduction and valence sides in **Fig. 2b**. $E_G$ is roughly ~300 mV at $\bar{D}$ = 2.5 V/nm.

We now consider the transport properties for the same device. **Figure 3a** shows the conductivity ($\sigma$) measured as a function of $V_{TG}$ for different $V_{BG}$ at 20 K. The conductivity at the Dirac point decreases with increasing $V_{BG}$, again indicating the band gap opening. The maximum $I_{on}/I_{off}$ is ~$3\times10^3$ for $\bar{D}$ = ~3.1 V/nm. **Figure 3b** shows the maximum mobility as a function of $\bar{D}$

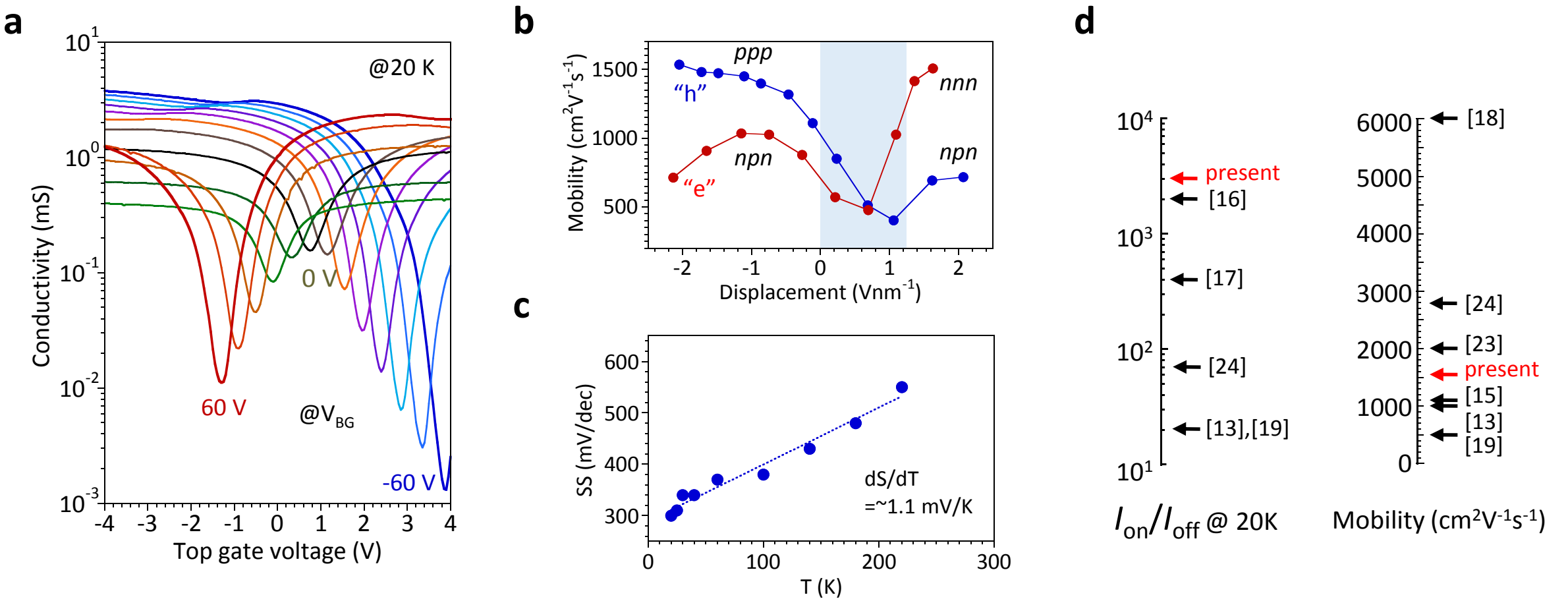

**Figure 3 | Transport characteristics.**

(a) $\sigma$ as a function of $V_{TG}$ for different $V_{BG}$ measured at 20 K. The drain current is on the order of 50 mA for the drain voltage of 100 mV. The leakage current is ~$10^7$ orders of magnitude lower than the drain current, as shown in supplementary **Fig. S1d**. (b) Maximum mobility as a function of $\bar{D}$ for different regions of *pnp*, *ppp*, *npn,* and *nnn*. The mobility at the hatched region (0 <$\bar{D}$<1.2) is relatively low because of the contribution of the ungated region between the top gate and source/drain electrodes (supplementary **Fig. S4**). (c) Subthreshold swing ($S$) as a function of temperature. The slope is nearly linear (d$S$/d$T$ = ~1.1 mV/K). (d) Comparison of $I_{on}/I_{off}$ and mobility with those reported previously for bilayer graphene. The temperature for $I_{on}/I_{off}$ data was fixed at 20 K. For the mobility, the temperature is not limited to 20 K, and both two-probe and four-probe data are shown here.

for different regions of *pnp*, *ppp*, and so on, whose positions are shown in supplementary **Fig. S4a**. In terms of the determination of carrier mobility using the Drude model ($\sigma = en\mu$), the carrier density ($n$) is estimated by the integration of the differential capacitance of $C'_{Total}$, that is, $n = 1/e \int_0^{V'_{TG}} C'_{Total}\, dV'_{TG}$, because of the large contribution of $C_Q$. Otherwise, $n$ is overestimated. However, the carrier density at the access region between the source (drain) and topgate is assumed to be consistent with that at the main channel region just below the topgate electrode for the mobility estimation. Shortening this access region by the self-alignment process[41] is important to extract the mobility more precisely and to improve the device performance.

Moreover, the subthreshold swing ($S$) is plotted as a function of temperature, as shown in **Fig. 3c**. The $S$ value (~600 mV/dec) extrapolated to room temperature is considerably larger than the theoretical lower limit of 60 meV/dec at room temperature[42], suggesting the large contribution of the gap states. **Figure 3d** compares the present mobility and $I_{on}/I_{off}$ at 20 K with previously reported data. All the data for $I_{on}/I_{off}$ are selected at 20 K from the literature, while the temperature for the mobility is not limited to 20 K. The mobility in the present study is not very high because it includes the contact resistance. In contrast, $I_{on}/I_{off}$ in the present study is quite high in spite of the direct deposition of high-*k* oxide on bilayer graphene without any organic buffer layer.

To determine $E_G$ from the viewpoint of the transport, we study the temperature dependence of $\sigma$ at different $\overline{\mathrm{D}}$ ranging from 0 to 3.1 V/nm. **Figure 4a** shows the temperature dependence of $\sigma$ as a function of $V_{TG}$–$V_{DP}$ at $V_{BG}$ = –60 V. So far, the temperature dependence of the conductivity in bilayer graphene is explained by the sum of three terms, the thermal activation (TA) at the high temperature region, nearest neighbor hopping (NNH) at the intermediate temperature range and variable range hopping (VRH) in a two-dimensional system of localized states at the low temperature region, as follows[17,18],

$$\sigma_{Total} = \sigma^0_{TA} exp\left[-\frac{E_G}{2k_BT}\right] + \sigma^0_{NNH} exp\left[-\frac{E_0}{k_BT}\right] + \sigma^0_{VRH} exp\left[-\left(\frac{T_0}{T}\right)^{1/3}\right],$$

where $\sigma^0_{TA}$, $\sigma^0_{NNH}$ and $\sigma^0_{VRH}$ are prefactors and $E_0$, $T_0$ and $k_B$ are the activation energy, the hopping energy and the Boltzmann constant. Since the lowest temperature in this study is 20 K, the contribution of VRH may not be observed. The conductivity at the Dirac point is plotted as a function of temperature and fitted by two combinations of TA + NNH and TA + VRH, as shown in **Figs. 4b and 4c**, respectively. All the conductivity data are well fitted with respect to $T^{-1}$ for TA + NNH, while the plot with respect to $T^{-1/3}$ does not show clear linear behavior below T = 100 K. The present data can be explained by TA + NNH for the temperature range above 20 K. This is consistent with the previous report where VRH was observed at below 5 K[13,17,18]. Extracted

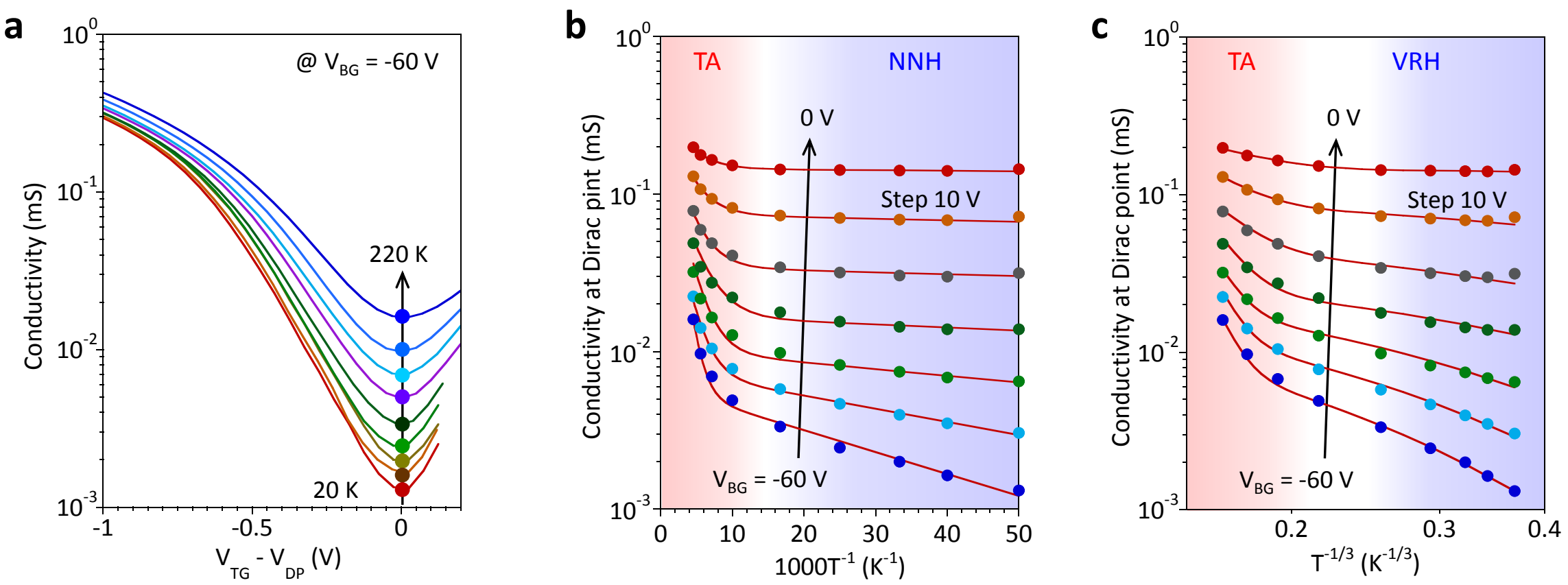

**Figure 4 | Temperature dependence of conductivity at the Dirac point.**

(a) Temperature dependence of $\sigma$ as a function of $V_{TG}$–$V_{DP}$ at $V_{BG}$ = –60 V. (b) and (c) $\sigma$ at the Dirac point as a function of temperature for different $V_{BG}$ ranging from –60 to 0 V. All the data are fitted by a sum of TA and NNH for (b) and by a sum of TA and VRH for (c).

$E_G$ and $E_0$ are plotted as a function of $\overline{D}$ in **Fig. 2c**, respectively, along with the results obtained in infrared absorption studies[14]. $E_G$ determined by *I*-*V* is reasonably consistent with the previous infrared absorption data, while $E_G$ determined by *C*-*V* is slightly larger. This difference is discussed later. Moreover, $E_0$ increases with $\overline{D}$ and reaches 2.8 meV at $\overline{D}$ = 3.1 V/nm, which is high compared with previous reports[18].

### Estimation of $D_{it}$ and $\tau_{it}$

To gain insight into the interface properties, the gap states in the electrostatically opened-band gap are quantitatively analyzed using the conductance method. The equivalent circuit of the device is shown in the left of **Fig. 5a**, where $R_{it}$ and $C_{it}$ are the resistance and capacitance associated with the interface traps, their product $C_{it}R_{it}$ is defined as $\tau_{it}$, and $R_s$ is the series resistance. It should be noted that $R_{it}$ and $C_{it}$ have been neglected due to roughly no electrical communication of carriers with trap sites at 1 MHz in **Fig. 2a**. Here, when this equivalent circuit is converted into $C_p$, in parallel with $G_p$, as shown in the right of **Fig. 5a**, the relation between $G_p$ and $D_{it}$ can be given as the following[35],

$$\frac{G_p}{\omega} = \frac{eD_{it}}{2\omega\tau_{it}} ln(1 + (\omega\tau_{it})^2), \quad (2)$$

where $\omega = 2\pi f$ (*f*: measured frequency). Finally, $D_{it}$ and $\tau_{it}$ can be obtained as $D_{it} = \left(G_p/\omega\right)_{max}/0.402e^2$ and $\tau_{it} = 1.98/2\pi f_0$, where $f_0$ is the frequency at $\left(G_p/\omega\right)_{max}$. The accurate estimation of $D_{it}$ is possible because $G_p/\omega$ does not include $C_Q$, which is the advantage of the conductance method. However, if there is leakage conductance ($G_{leak}$), indicated by the dotted line in the left of **Fig. 5a**, it is quite difficult to separate the two conductance contributions. The key issue to extract $D_{it}$ quantitatively is the fabrication of a highly reliable top gate insulator on bilayer graphene without any leakage current. This has been achieved by the high-pressure $O_2$ post-deposition annealing of $Y_2O_3$.

**Figure 5b** shows $G_p/\omega$ as a function of *f* for $E_F$ from -74 to -28 meV, where the band structure was fixed as constant at $\overline{D}$ = -1.5 V/nm, i.e., $E_G$ = ~150 meV. It should be emphasized that $G_p$ was measured under the constant band structure along the constant $\overline{D}$ line by controlling $V_{TG}$ and $V_{BG}$ at the same time. To prevent the contribution of a minority carrier response to the conductance, especially for small band gap semiconductors, the measurement temperature was lowered as much as possible, i.e., 20 K ($k_BT$ = ~1.7 meV). This quite low temperature is only possible in bilayer graphene because the majority carriers do not freeze due to the doping by the external gate control, unlike substitutionally doped semiconductors. The more detailed measurement conditions are described in the method section. Clear peaks were observed in the $G_p/\omega$-*f* relation, suggesting the detection of a trap-detrapping sequence of carriers with gap states. The $\tau_{it}$ value is evaluated to be ~4 μs. In **Fig. 2a**, it is assumed that the contribution of $C_{it}$ is roughly negligible in $C_{Total}$, measured at 1 MHz. According to this $\tau_{it}$ value, this assumption is roughly, but not completely, reasonable. Therefore, $E_G$ estimated by *C*-*V* may be slightly overestimated due to the contribution of $C_{it}$. **Figure 5c** shows $D_{it}$ as a function of $E_G$ for different $\overline{D}$. $D_{it}$ is in the range from the latter half of $10^{12}$ to $10^{13}$ $eV^{-1}cm^{-2}$. The energy dependence of $D_{it}$ is not so obvious. Although this value is slightly smaller than that for $MoS_2$[43], it is larger compared with those of Si[44]. The large amount of gap states at the interface of high-*k* oxide/bilayer graphene limits $I_{on}/I_{off}$ at present. The much lower $D_{it}$ could be achieved for dual-gate bilayer graphene heterostructure with *h*-BN.

In general, the conductance method is applied to the metal-oxide-semiconductor (MOS) capacitor, where the current is injected from the back side through the semiconductor substrate, whose series resistance, including back side contact, can be removed by measuring the impedance at the high-frequency limit (see Method). The interface trap conductance can be precisely measured. In contrast, in the present FET structure, the carriers are injected from the side of the channel through the source and drain[43,45]. The channel resistance may contribute to $G_p$ in the equivalent circuit. To avoid this, a device with a relatively short channel was used at the expense of the channel area, which increased $C_{para}$ in this study compared with typical value of ~0.2 $\mu Fcm^{-2}$ [36]. Supplementary **Figure S5b** compares $\tau_{it}$ with the channel time constant ($\tau_{ch} = R_{DP}C_{DP}$, where $R_{DP}$ and $C_{DP}$ are the measured resistance and capacitance at the Dirac point),

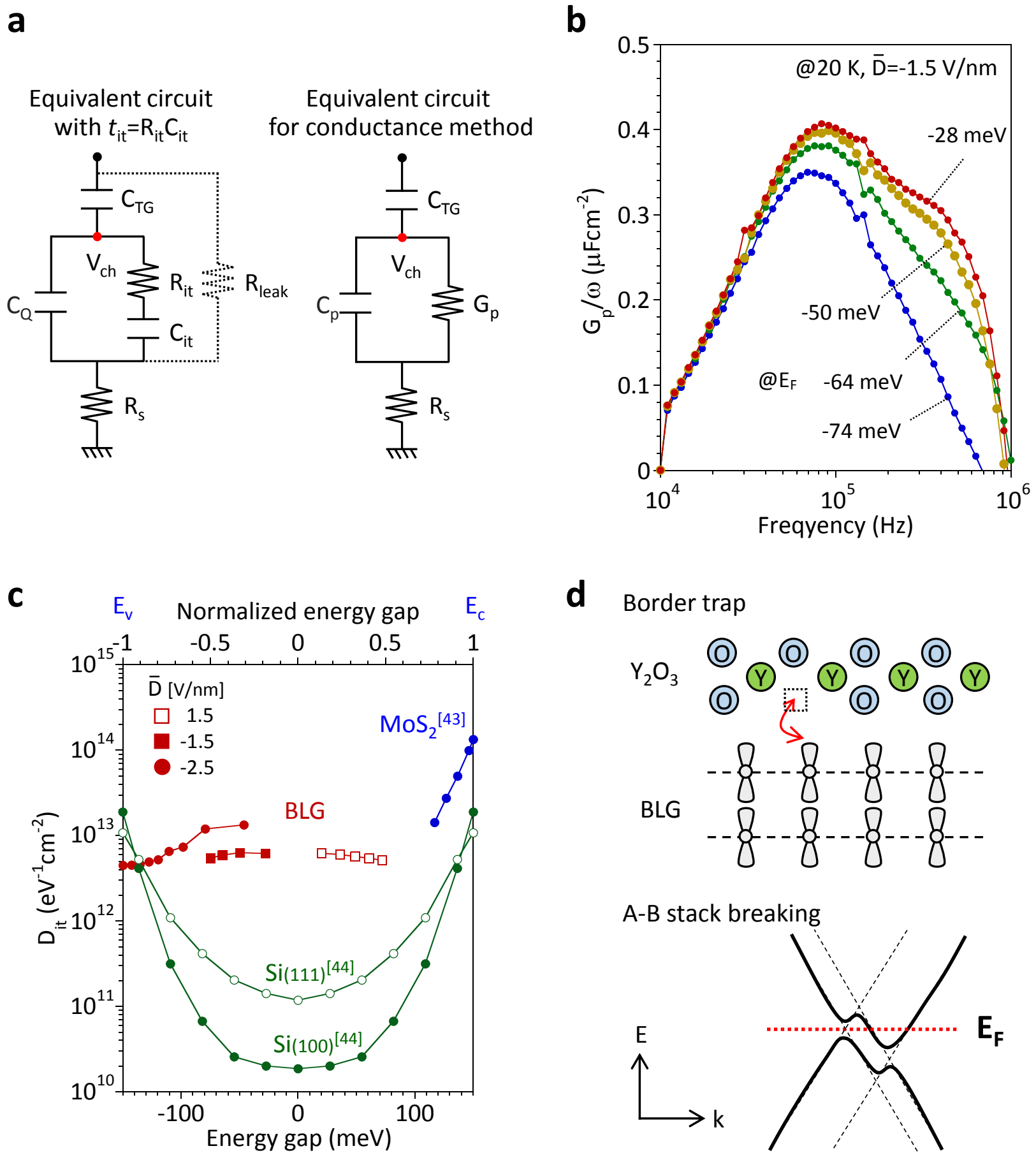

**Figure 5 | Characterization of gap states.**

(a) The equivalent circuits with $\tau_{it}$ = $R_{it}C_{it}$ (left) and for the conductance method (right). (b) $G_p/\omega$ as a function of $f$ at D̄ = -1.5 V/nm, i.e., the constant band structure with $E_G$ = ~150 meV. $E_F$ was changed from -74 to -28 meV. (c) $D_{it}$ of bilayer graphene (BLG) as a function of $E_G$ for different D̄. For comparison, $D_{it}$ for Si(100), Si(111)[44] and $MoS_2$[43] are also included as a function of normalized energy gap (top transverse axis). (d) Two possible origins of the gap states are shown. For the border trap (top), some defect sites in oxides, such as an oxygen vacancy (dotted box), work as trap sites. For A-B stack breaking (bottom), the band structure is not ideal but is like the dotted line. When the external electrical field is applied, the band gap is opened, like the solid line.

suggesting that $\tau_{it}$ is dominant. Therefore, the present $D_{it}$ measurement is reliable.

Finally, let us discuss the origin of the gap states. As mentioned before, no detectable defects are observed by the Raman D band. Therefore, the trap sites formed by defects are ignored in this discussion. Two possible origins are discussed here, as shown in **Fig. 5d**. One is the external origin, that is, the border trap[46]. The carriers in bilayer graphene electrically communicate with trap sites, such as the oxygen vacancy (dotted box) at the surface of oxides, because $\pi$ electrons stay on the surface of bilayer graphene. The other is the internal origin. The local breakdown of A-B stacking[47,48] results in the stacking of two monolayers[49], leading to the different band structure. In this case, $E_F$ stays within $E_G$ at some momentum space, while the real states exist at another momentum space, which may macroscopically act as gap states. Here, let us consider the dominant origin of the gap states. From the viewpoint of the time constant, the former origin will be a more time-consuming process than the latter because the rate limiting process is the tunneling between bilayer graphene and trap sites at the oxide surface[50]. On the other hand, if A-B stack breaking is the dominant origin of the gap states, $D_{it}$ should decrease with increasing $\overline{D}$, which is consistent with the observation in **Fig. 2b**. This, however, does not appear to be clear in the conductance measurement in **Fig. 5c** because other devices show almost no difference in $D_{it}$ for different $\overline{D}$. If $T_0$ in VRH is extracted at the low temperature region below 5 K, the $\overline{D}$ dependence of $D_{it}$ can be discussed more in detail because $T_0$ is inversely proportional to the density of states for the localized states[51]. Further detailed measurements are required to understand the $\overline{D}$ dependence of $D_{it}$. Although it is difficult to determine the dominant origin of the gap states at present, the quantitative recognition of the current status for $D_{it}$ in this study does provide the realistic comparison with other semiconductor channels in terms of the interface quality.

## Conclusions

We have succeeded in extracting critical information on the interface quality, that is, $D_{it}$ and $\tau_{it}$ for gap states. At the large displacement field of ~3 V/nm, they are in the range from the latter half of $10^{12}$ to $10^{13}$ $eV^{-1}cm^{-2}$ and ~4 μs, respectively. The large amount of gap states at the interface of high-*k* oxide/bilayer graphene limits the $I_{on}/I_{off}$ ratio at present. Many trials to characterize the interface properties for other 2D layered channels, as well as bilayer graphene, will help in understanding the origin of the gap states. The improvement of gap states below ~$10^{11}$ $eV^{-1}cm^{-2}$ is the first step for bilayer graphene devices to be a promising candidate for future nanoelectronic applications.

## Methods

The transport measurement of the device was performed using a Keysight B1500A semiconductor analyzer in a Nagase low-temperature prober in a vacuum of ~$10^{-5}$ Pa. The capacitance measurement and conductance method under the constant $\overline{D}$ conditions were performed by controlling both the Keysight 4980A LCR meter and the Keithley 2450 source meter for the back gate. The top gate electrode was connected to the high terminal, while the source and drain electrodes was connected to the low terminal. For the conductance measurement, $G_p$ should be measured accurately by reducing the contribution of $R_s$ in the equivalent circuit in the right side of **Fig. 5a**. Therefore, ozone treatment was performed for 5 min before the metal deposition to remove the resist residue at the source/drain contact[52], which reduces the contact resistance. Then, $R_s$ was estimated as follows. For the equivalent circuit in **Fig. 5a** (right), the impedance ($Z$) is given by $Z = \frac{1}{j\omega C_{TG}} + \frac{1}{j\omega C_p + G_p} + R_s = \left[\frac{G_p}{\omega^2 C_p^2 + G_p^2} + R_s\right] - j\left[\frac{1}{\omega C_{TG}} + \frac{\omega C_p}{\omega^2 C_p^2 + G_p^2}\right] = Z' - jZ''$. Therefore, $R_s$ at the accumulation side was estimated to be 29.4 kΩ by taking the high-frequency limit ($\omega \rightarrow \infty$), as shown in supplementary **Fig. S5a**. Finally, $G_p$ can be accurately estimated by removing $R_s$ in the equivalent circuit in the right side of **Fig. 5a**. For the $D_{it}$ measurement at the valence band side, $E_F$ is scanned along negative $\overline{D}$ (−1.5 and −3 V/nm). The access region modulated only by the back gate is always *p*-type (black line on the dashed-dotted vertical line at $V_{TG}$ = 0 V in **Fig. 1b**), from which a hole is injected into the main channel under the top gate to ensure majority carrier response, and vice versa for the conduction band side at the positive $\overline{D}$. Therefore, $D_{it}$ distributed throughout $E_G$ can be measured using the single FET device, unlike the MOS capacitor.

## Acknowledgments

The authors acknowledge Profs. Toriumi, Watanabe, Drs. Kasamatsu, Ando, and Minamitani and Mr. Mori for their helpful suggestions. We are grateful to Covalent Materials for kindly providing the Kish graphite. This research was partly supported by a Grant-in-Aid for Scientific Research on Innovative Areas and for Challenging Exploratory Research by the Ministry of Education, Culture, Sports, Science and Technology in Japan.

## Author contributions

K.N. conceived the experiment. K.K. and K.N. fabricated the devices, carried out the measurements and performed data analysis. K.N. wrote the manuscript.

## Additional information

Supplementary information is available in the online version of the paper. Reprint and permission information is available online at.www.nature.com/reprints. Correspondence and requests for materials should be addressed to K.N.

## Competing financial interests

The authors declare no competing financial interests.

Supplemental information

# Gap states analysis in electrostatically-opened band gap for bilayer graphene

Kaoru Kanayama[1], and Kosuke Nagashio[1,2*]
[1]Department of Materials Engineering, The University of Tokyo, Tokyo 113-8656, Japan
[2]PRESTO, Japan Science and Technology Agency (JST), Tokyo 113-8656, Japan
*nagashio@material.t.u-tokyo.ac.jp

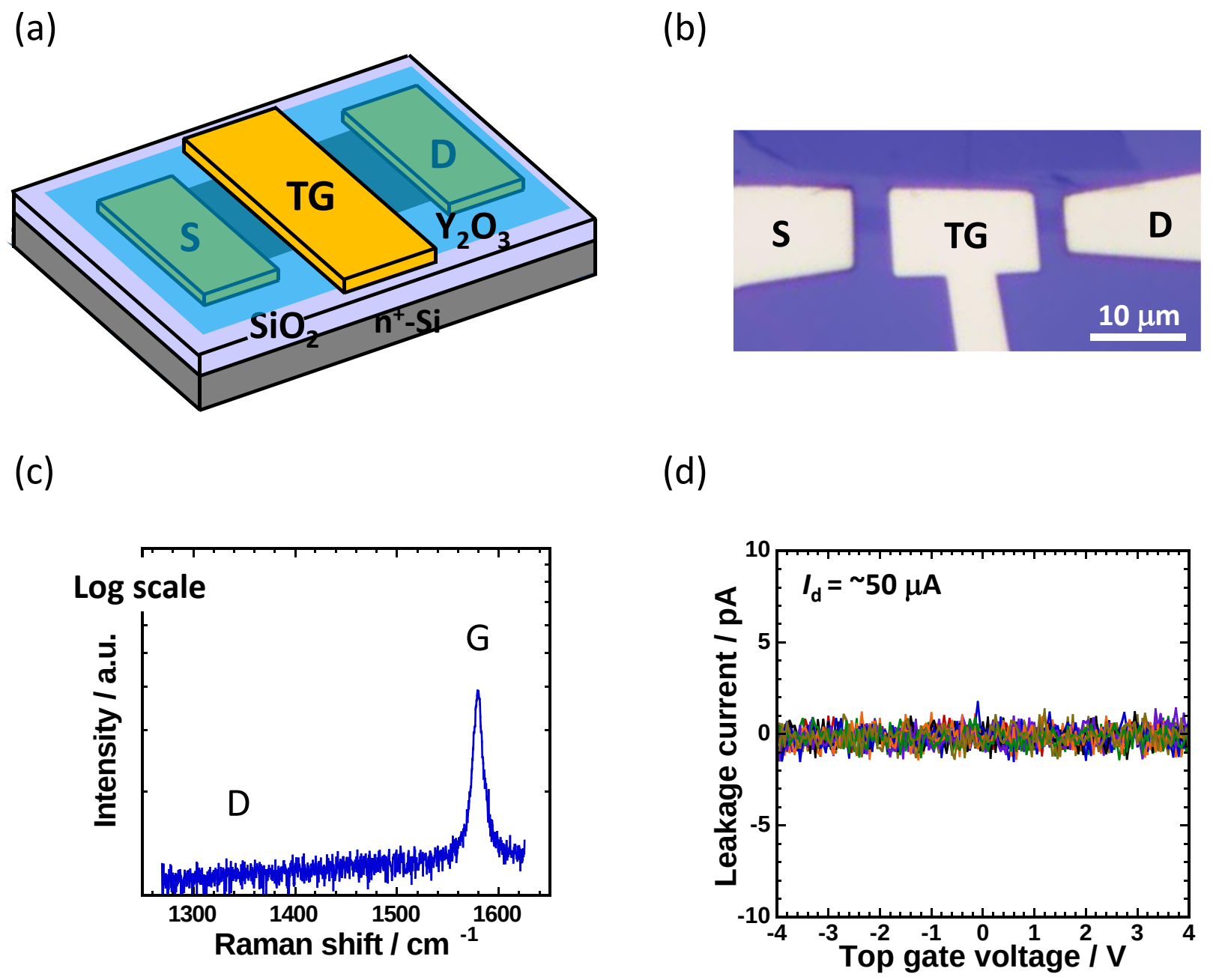

**Figure S1** (a) Schematic of a dual gate graphene bilayer FET. (b) Optical micrograph of the device. (c) Raman data obtained through $Y_2O_3$ top gate insulator after the high-pressure $O_2$ annealing. (d) Top gate leakage current obtained during $I_D$-$V_{TG}$ measurement. The drain current is roughly ~50 μA, while the top gate leakage current is a few pA for all different $V_{BG}$.

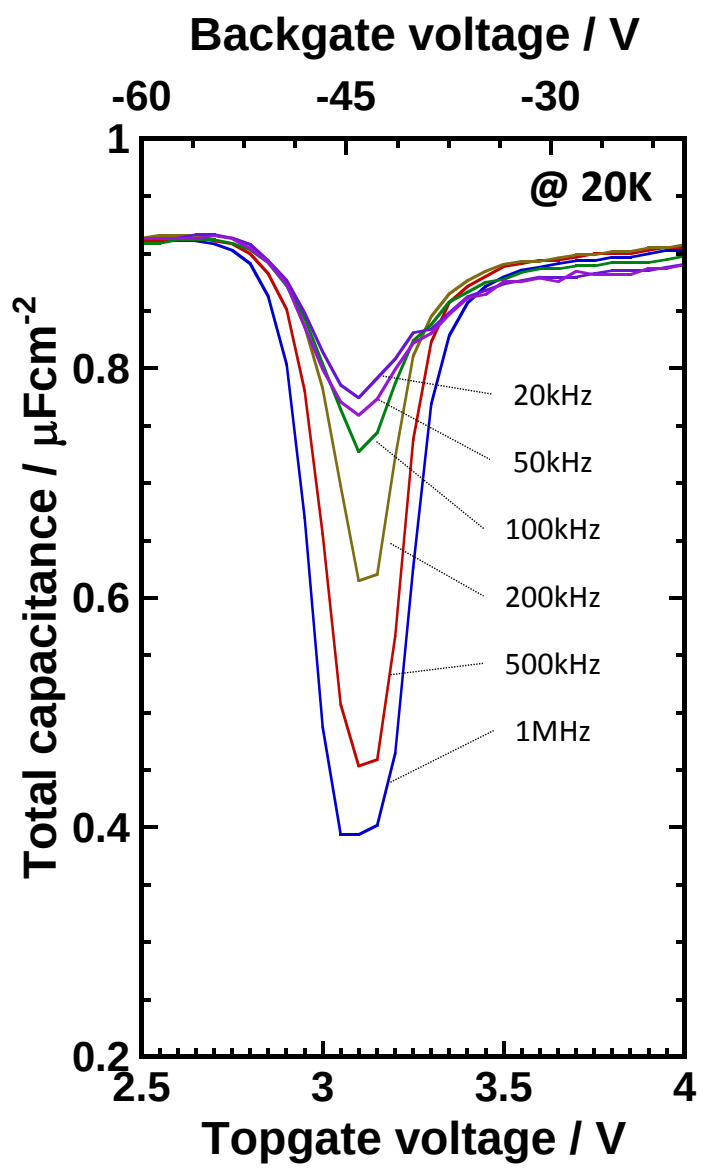

**Figure S2** Frequency dependence of $C_{\mathrm{Total}}$ at $\overline{\mathrm{D}}$=-2.5 V/nm, where $V_{\mathrm{TG}}$ and $V_{\mathrm{BG}}$ are changed at the same time, as shown in the upper and lower transverse axes. The source-drain-topgate device without bilayer graphene channel was fabricated just near the bilayer graphene FET device on the same wafer. The capacitance of the device without bilayer graphene shows some frequency dependence. This parasitic capacitance has been removed from this data.

Supplemental information

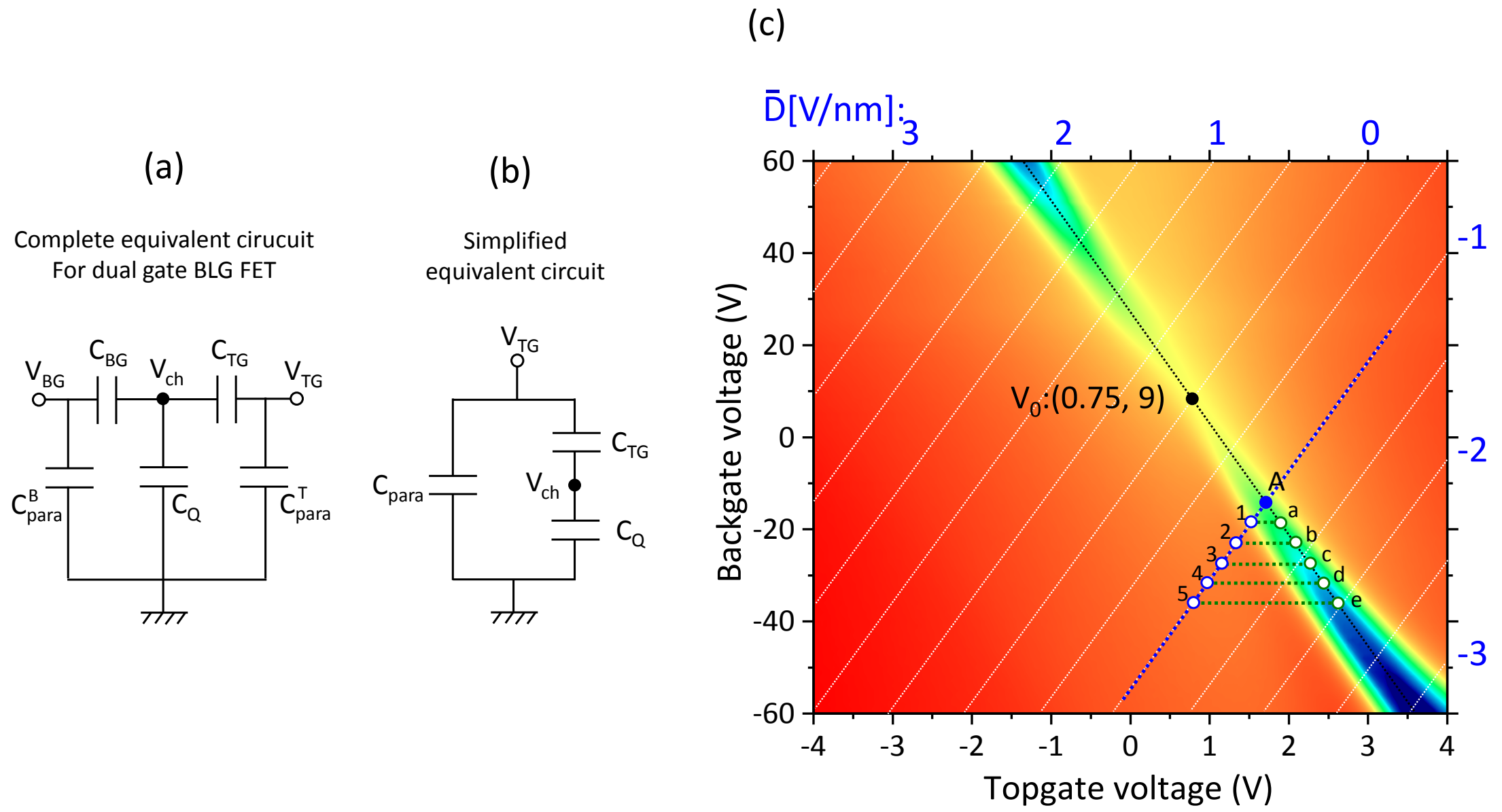

**Figure S3** In case of bilayer graphene, when $V_{TG}$ is applied under the constant $V_{BG}$ to change $E_F$, the band structure is also changed at the same time. Therefore, $C_Q$ should be extracted as a function of $E_F$ along the constant $\bar{D}$ lines to estimate $E_G$. The complete equivalent circuit for the dual gate bilayer graphene FET is shown in **Fig. S3a**. In this case, the determination of $V_{ch}$, i.e. $E_F$ (= $eV_{ch}$), is quite complicated. The equivalent circuit can be converted to **Fig. S3b**, where the contribution of $C_{BG}$ in $C_{Total}$ is involved through $C_Q$ and $V_{ch}$. Now, $V_{ch}$ can be calculated under the constant $V_{BG}$ by the following equation,

$$V_{ch} = V'_{TG} - \int_0^{V'_{TG}} C'_{Total}/C_{Y2O3}\, dV'_{TG}. \quad \text{[A]}$$

For the estimation of $V_{ch}$ for the position "1" along $\bar{D}$ = −1 V/nm in **Fig. S3c**, $C'_{Total}$ (differential capacitance) should be integrated from the position "a" to the position "1" along the constant $V_{BG}$ line. This is equivalent with the integration of $C'_{Total}$ from the position "A" to the position "1" along the constant $\bar{D}$, because the final position "1" is the same. Altough the contribution of $C_{BG}$ is not explicitly apparent in the equation [A], the change in the band strcuture and $E_F$ by $C_{BG}$ are implicitly included in $C'_{Total}$. To obtain $C_Q$ as a function of $E_F$, this calculation was iterated "1", "2", "3", "4", and "5" along the constant $\bar{D}$.

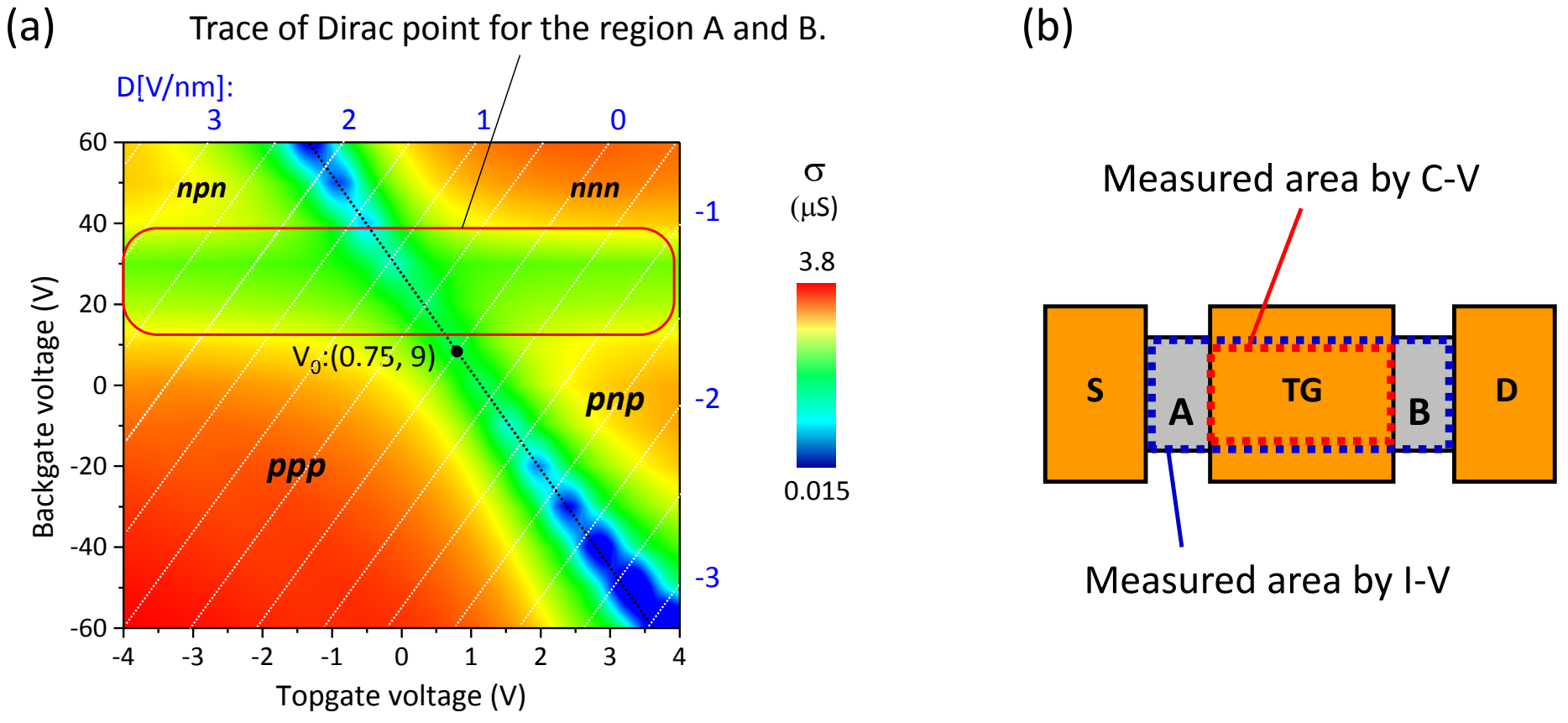

**Figure S4** (a) Counter plot of $\sigma$, showing the transverse constant $\sigma$ region. The access region between source(drain) and topgate (A and B in **Fig. S4b**) is only modulated by the backgate. The transverse constant $\sigma$ region is formed due to the trace of Dirac point for A and B regions. Therefore, the mobility at the hatched region ($0 < \overline{D} < 1.2$) in **Fig. 3b** is relatively low.

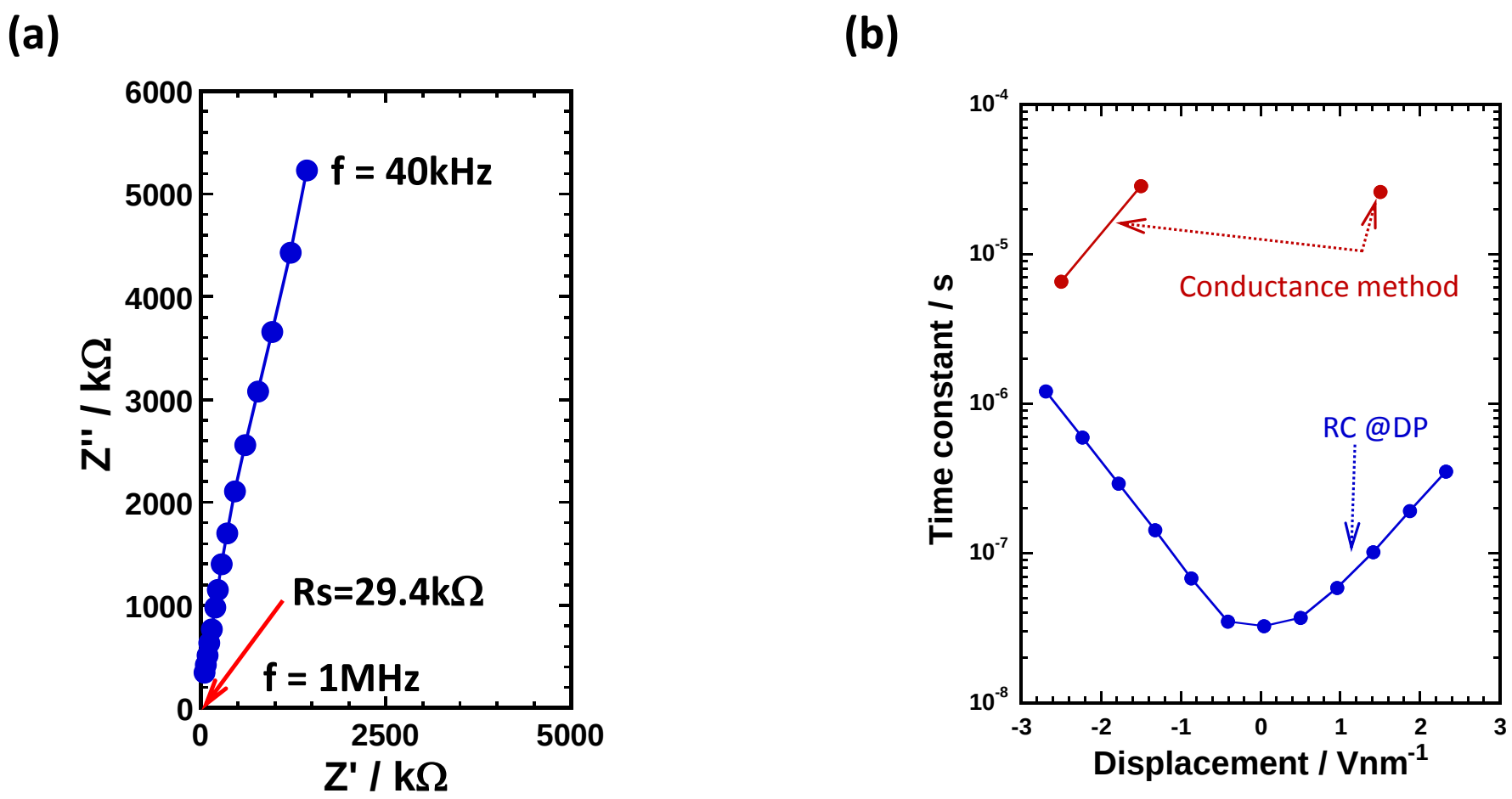

**Figure S5** (a) Z'' as a function of Z'. $R_s$ is estimated to be 29.4 kΩ. (b) Time constant as a function of $\overline{D}$ for $\tau_{it}$ and $\tau_{ch}$, where $\tau_{ch} = R_{DP}C_{DP}$, where $R_{DP}$ and $C_{DP}$ are the measured resistance and capacitance at the Dirac point.